# Graphene nano-ribbon waveguides

S. He*, X. Zhang, and Y. He

*JORCEP, COER, Zhejiang University, Hangzhou 310058, China*

Abstract

Graphene as a one-atom-thick platform for infrared metamaterial plays an important role in optical science and engineering. Here we study the unique properties of some plasmonic waveguides based on graphene nano-ribbon. It is found that a graphene ribbon of finite width leads to the occurrence of coupled edge mode. The single-mode region of a single freestanding graphene ribbon is identified at a fixed frequency of 30 THz. A low-loss waveguide structure, consisting of a graphene layer, a silica buffer layer and silicon substrate is proposed to reduce the propagation loss and obtain a high figure of merit for future integration of waveguide devices. Furthermore, two coupled ribbon configurations, namely, side-side coupling and top-bottom coupling, are investigated. As a device example, a nano-ring cavity of ultra-small size is designed,

Since graphene was experimentally made for the first time from graphite in 2004, this newly available material, which is a single layer of carbon atoms, has attracted much attention due to its unique properties[1,2]. It can serve as a platform for metamaterials and can support surface-plasmon polaritons (SPPs) at infrared or THz frequencies[3-7,11]. Compared with noble metals, e.g., Au and Ag, which are widely regarded as the best platform for SPPs, graphene can be tuned flexibly via electrical gating or chemical doping[3]. It can support both TM and TE modes[4] and has strongly enhanced light–matter interactions[7]. Meanwhile, graphene exhibits relatively low Ohmic losses. Some SPP on a graphene ribbon of finite width has been studied[11]. In this paper, low-loss plasmonic waveguides of ultra-small mode area based on some graphene nano-ribbon structures are proposed and studied.

As a unique property of graphene, the complex conductivity of a graphene layer can be tuned flexibly by controlling the chemical potential (via the applied electric field), chemical doping or ground plane evenness[3]. This way graphene can behave as a metal or dielectric and support both TM and TE waveguide modes. The graphene considered throughout the present paper has a chemical potential $\mu_c$= 0.15 eV, T= 3 K and scattering rate $\Gamma$= 0.43 eV to achieve a conductivity value $\sigma_g$= 0.0009+i0.0765 ms, which is capable of supporting TM SPP surface mode [4,5,7,9] at 30 THz. The conductivity value $\sigma_g$ is derived using the Kubo formula model[3,8]. The dispersion relations of TM SPP surface mode is expressed as $\beta^2=k_0^2[1-(2/\eta_0\sigma_g)^2]$, where $\beta$ and $k_0$ are the wave numbers in the wave guide and the free space, respectively, and $\eta_0$ is the intrinsic impedance of free space. Thus, we get $\beta = 69.34k_0$ for an infinitely extended graphene sheet.

As a 2D metamaterial, graphene with infinitely-small thickness cannot be directly incorporated into the numerical simulation using a conventional electromagnetic software such as COMSOL or CST. This difficulty can be overcome by taking the graphene layer as an ultra-thin material with an effective thickness $\Delta$ and effective bulk conductivity[3] $\varepsilon_{g,eq} = -\sigma_{g,i}/\omega\Delta+\varepsilon_0+i\sigma_{g,r}/\omega\Delta$, where $\sigma_{g,r}$ and $\sigma_{g,i}$ are the real part and imaginary part, respectively, of $\sigma_g$. In all our simulations, the thickness of graphene $\Delta$ is assumed to be 0.4nm (very close to the typical value of 0.34nm measured using the interlayer space of graphite [10]).

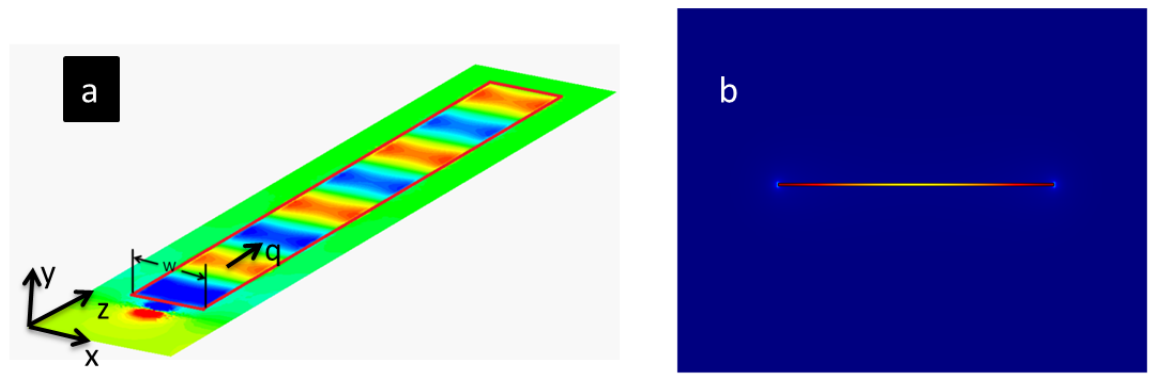

**Figure 1 | the structure and the energy density of the waveguide mode in a single freestanding graphene ribbon.** (**a**) Plasmonic mode supported by a single freestanding graphene ribbon. The boundary of the graphene ribbon is indicated by the red line. The width of the graphene ribbon is $w = 40$ nm. A discrete port is used for the excitation of the waveguide mode. (**b**) The distribution of the energy density of the waveguide mode. One clearly sees that the energy is tightly confined inside the graphene ribbon.

The waveguide mode of a single freestanding graphene ribbon is studied first. The 3D

propagation of a guided wave on a typical graphene ribbon with width $w = 40$ nm is first shown in Fig. 1a to give the readers an intuitive idea on the mode confinement in both the lateral direction and the propagation direction. A discrete port is placed in front of the graphene ribbon to excite the guided mode. Although the length along the propagation direction is only 350 nm, more than 4 harmonic oscillations are supported, indicating the existence of waveguide mode with an ultra-large wave number. Meanwhile, the waveguide mode is also tightly confined in the lateral direction, as can be seen from the rapid decay of the optical field in the surrounding air. Since an ultra-small waveguide with some tightly confined energy is highly desirable for nanoscale photonic integration, it is worthwhile to study the energy profile of the waveguide mode. The electromagnetic energy density $W(x, y)$ of the corresponding guided mode in Fig. 1a is shown in Fig. 1b. Here $W(x, y)$ is calculated by $W = 0.5(\varepsilon_{eff}\varepsilon_0|E|^2 + \mu_0|H|^2)$, where $\varepsilon_{eff}$ is the effective permittivity and $\varepsilon_{eff} = \partial(\omega\varepsilon)/\partial\omega$. Based on the dispersion model, our calculation gives $\varepsilon_{eff}=169.45$ when the thickness $\Delta$ *of the* graphene layer is 0.4nm and $\varepsilon_{eff,air}= 1$. In Fig. 1b, one sees that the electromagnetic energy is tightly confined inside the graphene due to the large magnitude of $\varepsilon_{eff,g}$. It is also noted that a significant amount of energy is concentrated on the edges of the graphene ribbon.

The evolution of the waveguide mode in a graphene ribbon of finite width has already been studied in ref.11. It is found that a new waveguide mode will appear due to the presence of the graphene edge. Here we want to design some graphene waveguides with extremely modal spotsizes at a fixed frequency.

The dependence of the effective refractive index on the width of a freestanding graphene ribbon is shown in Fig. 2a. It is found that the number of modes will decrease as the width decreases. When the ribbon width is large (> 200 nm), the effective index of the waveguide mode arising from the surface plasmon mode in an infinitely extended two-dimensional graphene (2DGSP) (called mode 3) would decrease as width $w$ gets smaller. These modes will be cutoff when $w$ decreases further, leaving only two modes (namely, mode 1 and mode 2) originating from the hybridization of the edge graphene surface plasmon (EGSP) modes supported by a semi-infinite graphene. As shown in Fig. 2b, $E_y$ component of mode 1 is symmetric with respect to the y axis, and $E_y$ component of mode 2 is anti-symmetric with respect to the y axis. Mode 1 has a higher refractive index than mode 2. As the width further shrinks, these two modes will continue splitting, and finally mode 2 is cutoff when $w < 50$ nm. Eventually, the graphene ribbon can operate at single-mode region (red shaded region in Fig. 2a) if its width is small enough. Interestingly, this remaining mode will not be cutoff even if the ribbon width further shrinks. The mode area is defined as $A_{eff}=\iint W(x,y)dxdy/W_{max}$, where $W_{max}$ is the maximum of the energy density of the whole waveguide cross section. The mode area is displayed in Fig. 2c as a function of ribbon width and decreases considerably as the width gets small. That is because the optical energy with the graphene ribbon dominates the whole energy, so a waveguide with smaller width can lead to a smaller mode area. In our calculation, an extremely small mode area of $1.3 \times 10^{-7} \lambda_0^2$ (fig.3) is obtained at $w = 20$ nm, which is the smallest among all the reported subwavelength waveguide (to the best of our knowledge).

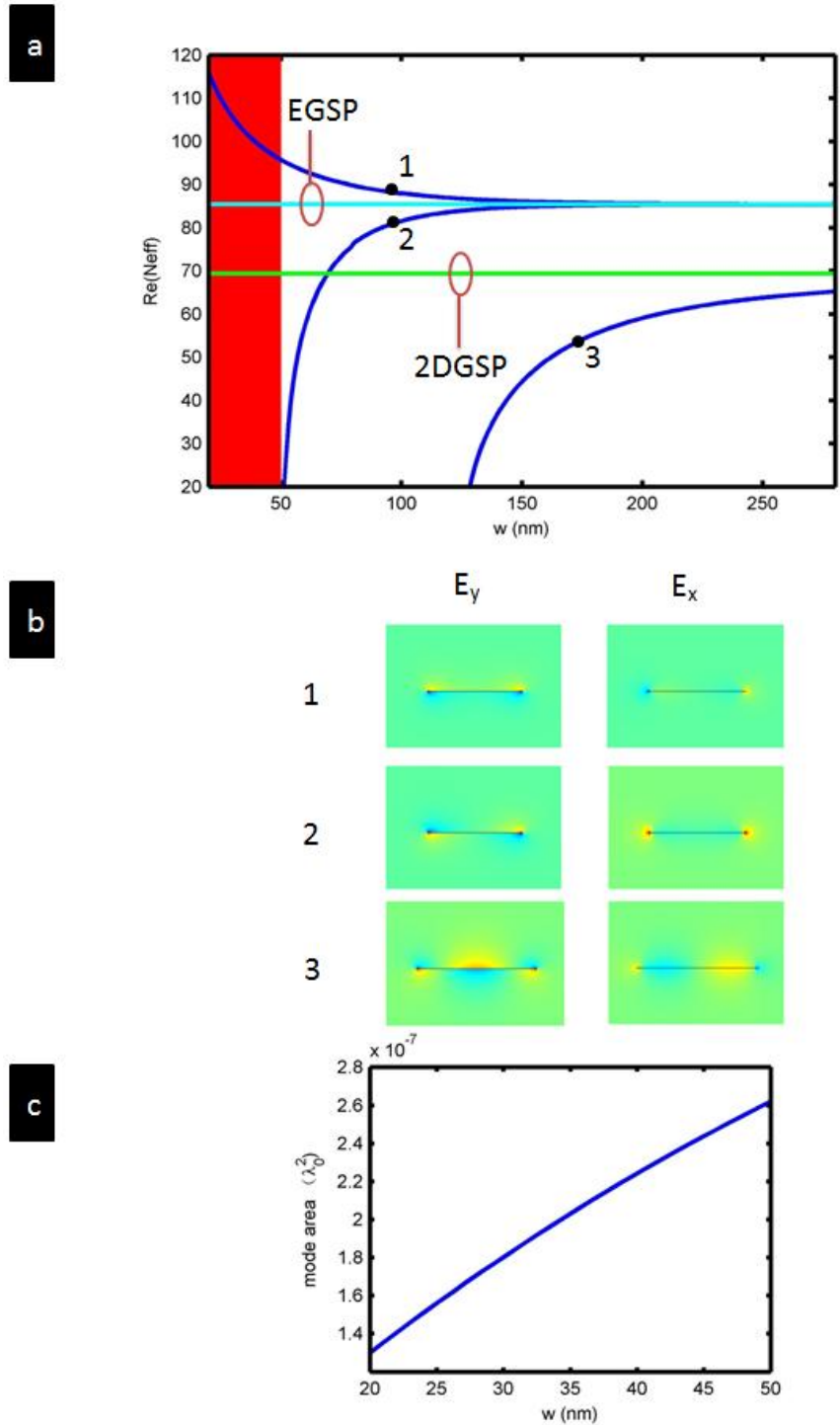

**Figure 2 | The effective refractive index, electric field and the mode area of the single freestanding graphene ribbon waveguide.** (a) The dependence of the effective refractive index on the width of the graphene ribbon. The dark blue lines show the variation of three SPP waveguide modes when the width of the ribbon varies. The green line marked as "2DGSP" is for the SPP waveguide mode in an infinitely-extended graphene sheet. The light blue line marked as "EGSP" is for the SPP waveguide mode in a semi-infinite graphene sheet[11]. We find that modes 1 and 2 originate from the hybridization of EGSPs. Mode 1 has an even parity of $E_y$ with respect to the ribbon axis, and its effective index $n_{eff} > n_{eff}^{EGSP}$. In contrast, mode 2 has an odd parity of $E_y$ with respect to the ribbon axis,

and its effective index $n_{eff} < n_{eff}^{EGSP}$. Mode 2 will be cut off if the ribbon is narrow enough. Thus, there is a single-mode region (red shaded region in (**a**)) for small width $w$. The effective index $n_{eff}$ of mode 3 will increase as $w$ increases and finally approach to $n_{eff}^{2DGSP}$ [which is not shown in (**a**)]. (**b**) The electric fields of the three modes. (**c**) The mode area ($A_{eff}/\lambda_0^2$) of the SPP waveguide in a freestanding graphene ribbon when width $w$ is in the single-mode region.

Next, we propose to reduce the propagation loss by buffering the graphene ribbon with a silica layer on a silicon wafer, as shown in Fig. 3a. $h_{SiO2}$ is the height of the SiO$_2$ layer, and $h_{Si}$ is the height of the Si layer. For this structure, we choose $w$=20 nm, $h_{Si}$ =20 nm, $\varepsilon_{si}$= 11.9, and $\varepsilon_{SiO2}$= 2.09. If there is no SiO$_2$, the effective index is $n_{eff}$= 466.3+4.7i. Accordingly, the guided wavelength $\lambda_{spp}$= $\lambda_0$/Re($n_{eff}$)= 21.4 nm. The propagation length $L_m$ defined as 1/Im($\beta$), where Im($\beta$)=Im($n_{eff}$)$k_0$, and $k_0 = \omega\sqrt{\mu_0\varepsilon_0}$. Thus, the propagation length is 338 nm, and the figure of merit (FOM) [which is defined as the ratio of Re($n_{eff}$) to Im($n_{eff}$)[12]] of this waveguide mode is 99. When a thin layer of SiO$_2$ exists, a strong electric field is induced within the SiO$_2$ layer due to the high index contrast between Si and SiO$_2$ (as required by the continuity of the normal displacement field components). Consequently, the percentage of the optical energy inside the graphene layer decreases, which leads to a reduced propagation loss since the optical loss is entirely due to the damping inside the graphene. Taking $h_{SiO2}$ = 5 nm as an example, we have $n_{eff}$= 180.9+1.3i, $L_m$ = 1256 nm, and $\lambda_{spp}$ = 55.3 nm. Compared with the case when the SiO$_2$ layer is absent or a freestanding graphene ribbon, significant loss reduction is achieved. Although the Im($n_{eff}$) of our slot structure is larger than that of the freestanding graphene ribbon (indicating short propagation length), the Re($n_{eff}$) is also greatly enhanced due to the presence of high-index substrate. It turns out that our slot waveguide has a larger FOM compared to that of the freestanding case. Si has a high index and thus will pull the light away from the graphene to the SiO2 buffer layer. Since more energy is located in the region of the low refractive index SiO$_2$ layer, the effective index of the waveguide mode Re($n_{eff}$) decreases at the same time. A good effect is that the FOM of such a graphene ribbon waveguide increases to FOM = 142.9 (much larger than that for the structure without SiO$_2$,). By changing $h_{SiO2}$, the effective index Re($n_{eff}$) varies from 160 to 400 (see Fig. 3e). It is worth noticing that there is a maximum FOM when the height of SiO$_2$ varies. By comparing with the structure without SiO$_2$, the maximum figure of merit Re($n_{eff}$)/Im($n_{eff}$)≈145 is obtained when $h_{SiO2}$≈ 3.0 nm, which is 50 % larger than that for the structure without SiO$_2$. For the structure with a maximum FOM, we have $n_{eff}$ = 204.4+1.4i (i.e., the effective wavelength of the guided wave is $\lambda_{spp}$=48.9nm), and $L_m$ = 1126 nm. In comparison, for a freestanding graphene waveguide with the same width, we have $n_{eff}$= 117.5+0.9i, $\lambda_{spp}$=89.1nm and $L_m$ = 1785nm. Since the effective wavelength in the waveguide on Si substrate is only about half of that in the freestanding graphene waveguide, more propagation cycles can be supported (before the light is very much attenuated) in our current design. Typically the size of an optical device is on the order of $\lambda_{spp}$. The graphene ribbon (with a silica buffer layer) on a silicon wafer provides a much smaller $\lambda_{spp}$ than the freestanding graphene with the same width. Here, high-index Si pulls the light away from the lossy graphene to the lossless SiO$_2$ layer and helps to improve FOM.

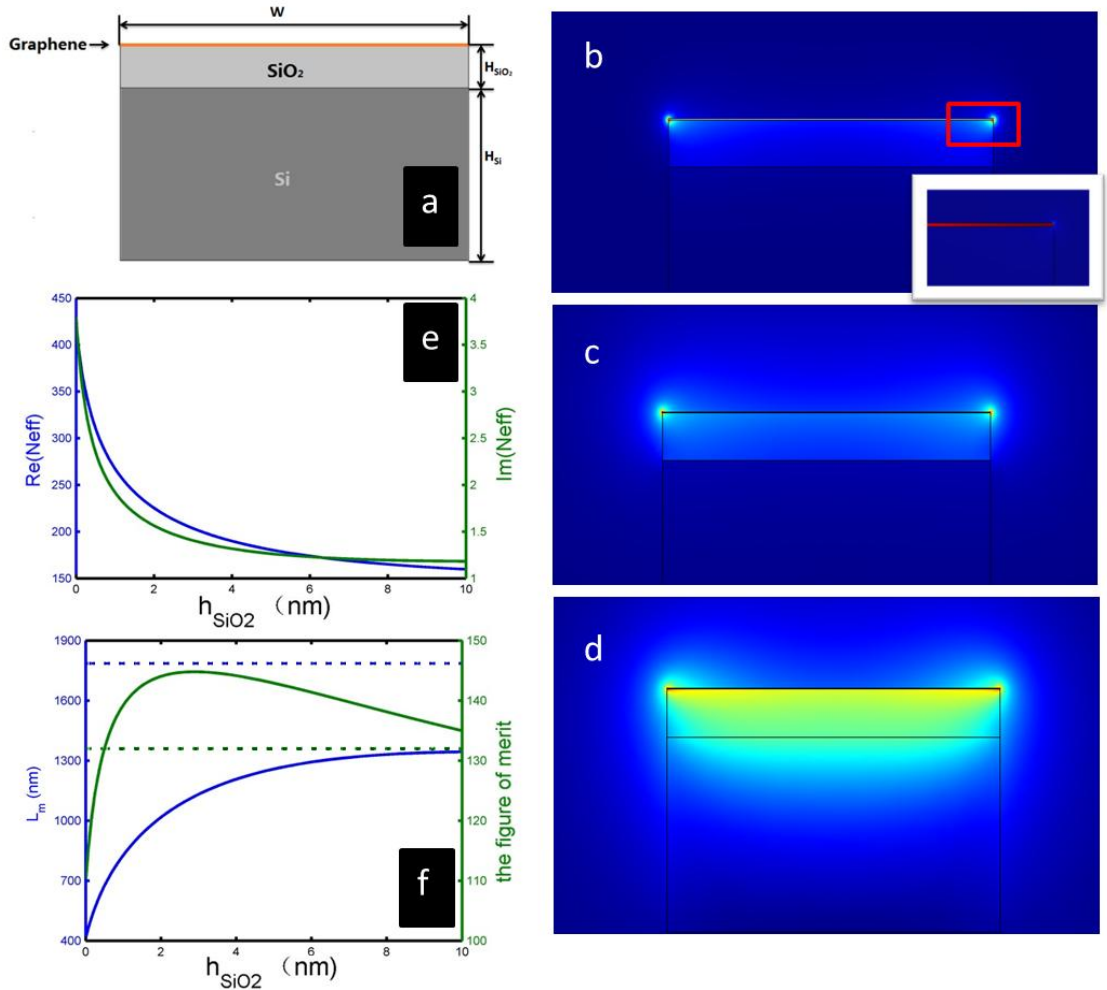

**Figure 3 | A low-low waveguide structure.** (**a**) A low-low waveguide structure: a graphene buffered by a silica layer on silicon. (**b**) The distribution of the energy density in the waveguide structure. Most of the energy is in the graphene and silica layer. Inset of (**b**) shows the energy field at the edge of the graphene. It shows the energy is tightly confined in the graphene and $W_{max}$ is located at the edge of the graphene. (**c**) and (**d**) are the distributions of the electric field and the magnetic field in the waveguide structure. In (**c**), it is found that the electric field is strong in the graphene layer and the silica (the electric field in silicon is very weak), especially at the corners of the graphene ribbon. In (d), the magnetic field is strong in all the three layers. (**e**) shows the effective index $n_{eff}$ of the low-loss waveguide structure when $w$ varies. The blue solid line is Re($n_{eff}$) and the green dashed line is Im($n_{eff}$). (**f**) The solid green line is the FOM of the SPP waveguide for this structure with different $h_{SiO2}$. The dashed green line is the FOM for a freestanding graphene waveguide with the same width (20 nm). As $w$ gets smaller, the FOM will first increase for $h_{SiO2}>3.0$ nm and then decreases quickly for $h_{SiO2}<3.0$ nm, leading to a maximum FOM = 145 and it is bigger than the FOM for a freestanding graphene waveguide with the same width. The solid blue line is the propagation length for this structure with different $h_{SiO2}$, and the dashed blue line is the propagation for a freestanding graphene waveguide with the same width. The propagation length increases as $h_{SiO2}$ increases.

Finally, we would like to study the mode properties of a graphene waveguide formed by two graphene ribbons. Similar to the case of mode coupling in a conventional MIM (metal-insulator-metal) or IMI (insulator-metal-insulator) waveguide, it is expected that the coupling of the nano-ribbon waveguide can lead to further mode splitting. Thus, the waveguide constructed by two identical graphene ribbons with a nanometer gap is investigated in this work. Two kinds of coupling configurations, namely, the side-side coupling (Fig. 4a) and top-bottom coupling (Fig. 4b), are studied.

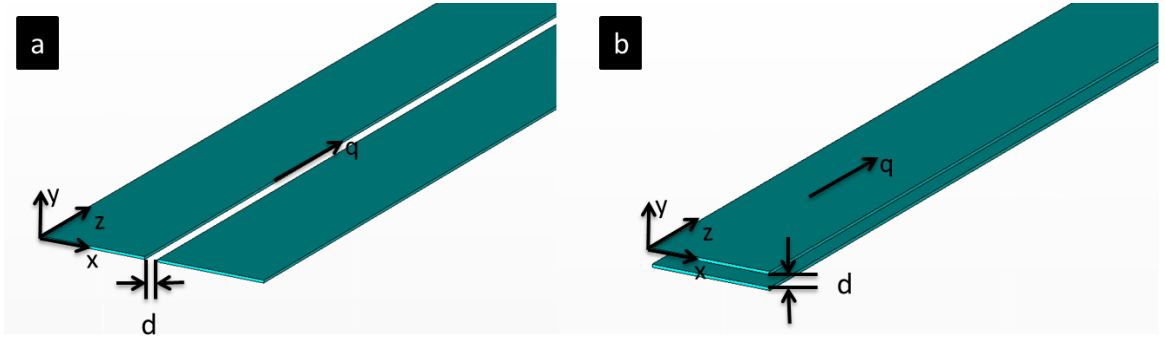

**Figure 4 | The structure of two models.** (**a**) is the side-side model and (**b**) is the top-bottom model. $d$ is the gap size between the two ribbons. The wave propagates in the direction of the vector $q$.

For the side-side configuration, $d$ is the distance between the two graphene ribbons, and the width of graphene ribbons is chosen to be 20 nm. If the gap between the ribbons is small enough, strong hybridization of the waveguide modes will occur, leading to the presence of both a symmetric mode (Fig. 5a) and an anti-symmetric mode (Fig. 5b). It is found that the symmetric mode can squeeze the optical energy effectively into the gap between the two ribbons, but the anti-symmetric mode only slightly modifies the profile of the optical energy density. For the symmetric mode, $n_{eff}$ and the figure of merit increase when $d$ is reduced (Fig. 6a). The waveguide mode area depends critically on the gap size (Fig. 6b). A smaller gap will lead to a smaller mode area. The mode area is extremely small, only about $10^{-7} \lambda_0^2$, which is an order smaller than the smallest mode area of any waveguide that has ever been reported in the literature.[13]

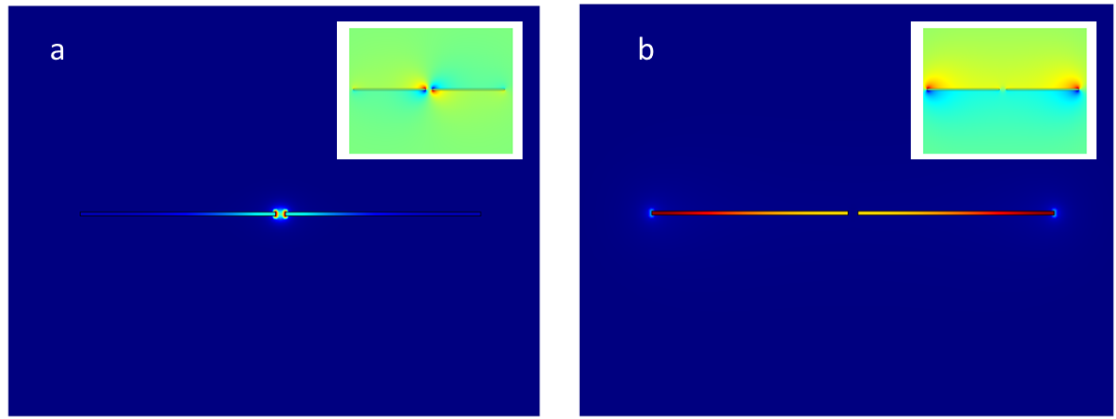

**Figure 5 | The energy density of SPP waveguide modes in the side-side configuration.** (**a**) is for the symmetric mode and (**b**) is for the anti-symmetric mode. In (**a**), the energy is mainly in the gap between the two ribbons, and in (**b**) the highest energy density is at the left and right corners. The insets in (**a**) and (**b**) show the distribution of the electric field of the modes.

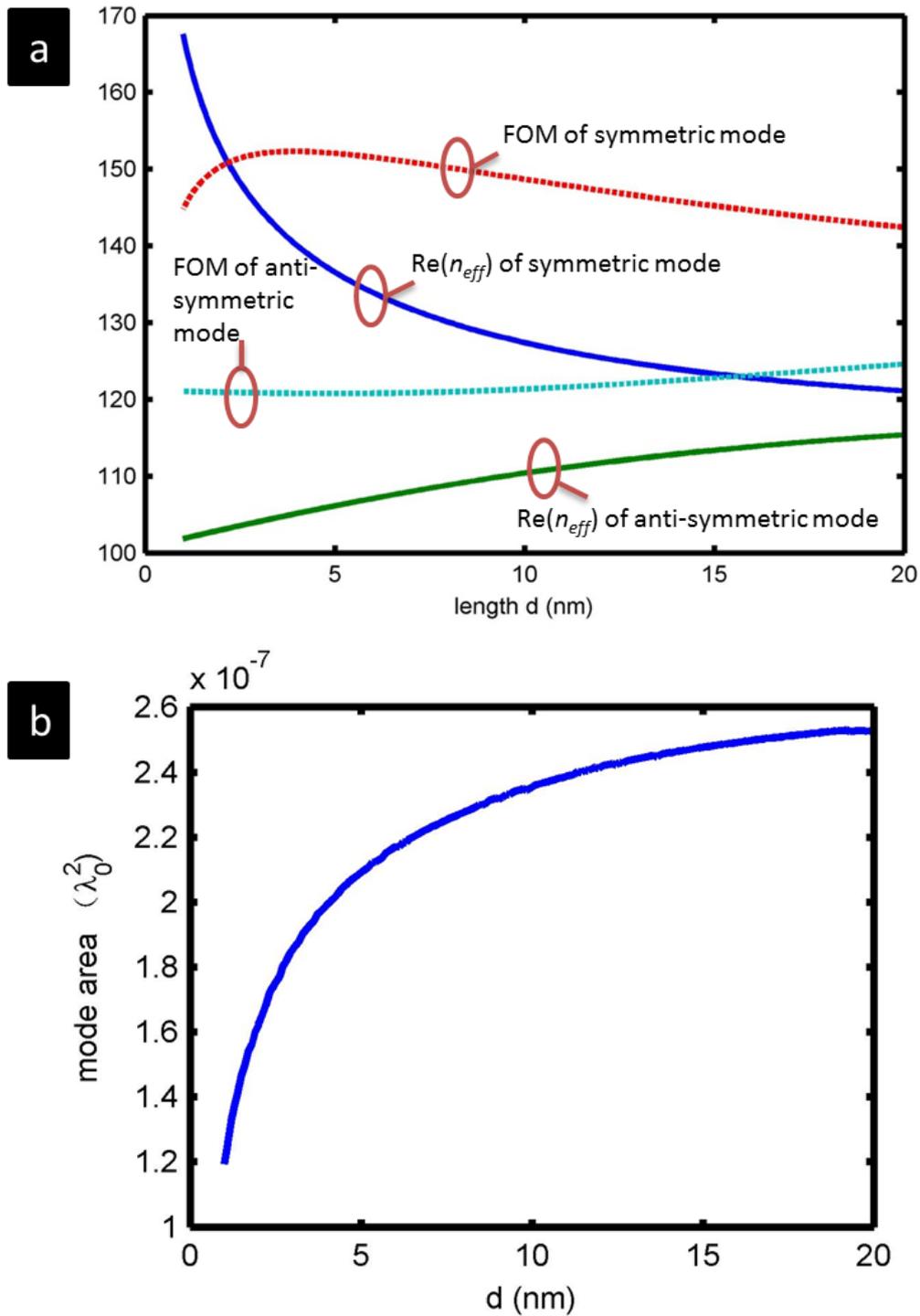

**Figure 6 | The side-side configuration. (a)**, the dashed blue solid line is for Re($n_{eff}$) of the symmetric mode, the green solid line is for Re($n_{eff}$) of the anti-symmetric mode, the dashed red line is the figure of merit of the symmetric mode, and the light blue dashed line is the figure of merit of the anti-symmetric mode. **(b)** The mode area of the symmetric mode. The mode area decreases rapidly as the gap $d$ decreases.

For the top-bottom model, $d$ is the vertical distance between the two ribbons, and the width of the graphene ribbons we use is also 20 nm. Again, there are both a symmetric mode and an anti-symmetric mode due to the presence of mode coupling. Fig. 7 shows the distributions of energy, electric field and magnetic field of the two modes. For the symmetric mode, more energy is confined in the area between the two ribbons. For the anti-symmetric mode, the field is mainly located at the top and bottom regions (the energy density, however, is still concentrated within the graphene ribbon due to the large effective permittivity $\varepsilon_{eff}$). Fig. 8 displays Re($n_{eff}$) of the two modes. The effective refractive index of the waveguide mode depends critically on distance $d$. If the quantum effect is not taken into account, Re($n_{eff}$) can be extremely large, reaching a value of up to 1000 when $d$ is smaller than 1 nm. The figure of merit is also a function of distance $d$, and can be more than 180. However, the mode area remains more or less the same when $d$ changes due to the large amount of energy in graphene.

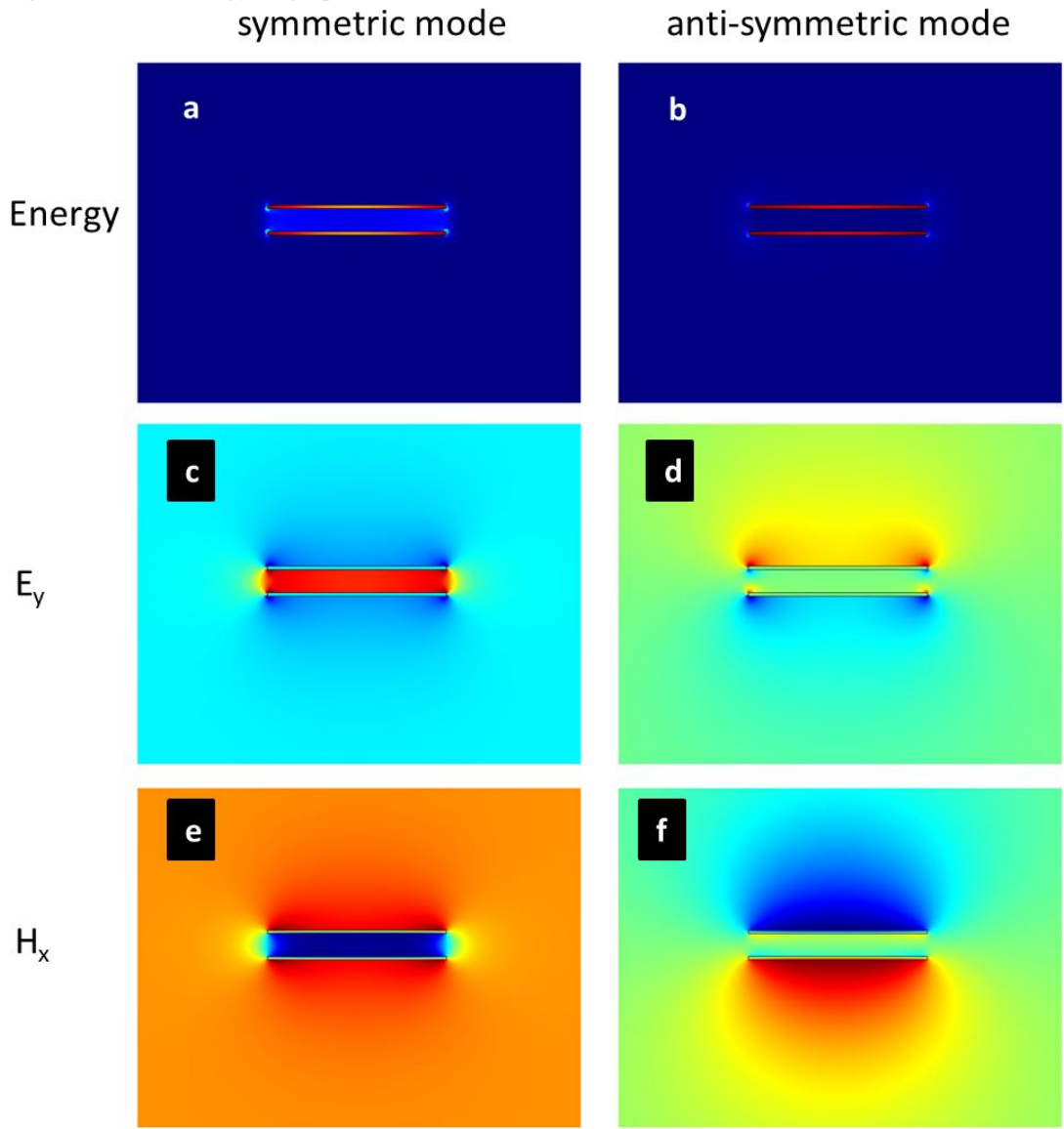

**Figure 7 | The distribution of the energy intensity, electric field, and magnetic field for the two modes in the top-bottom configuration.**

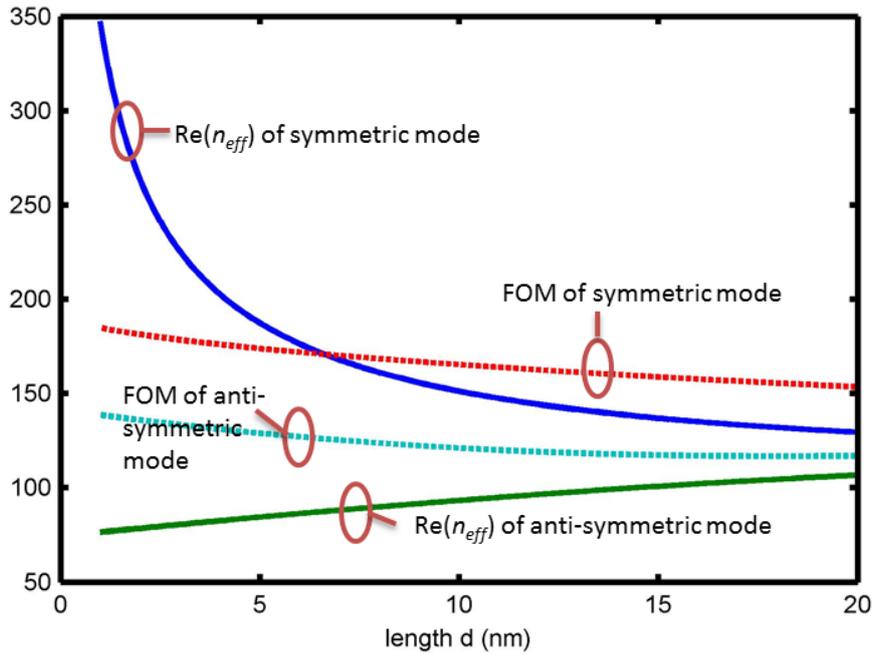

**Figure 8 | The top-bottom configuration.** The blue solid line is for Re($n_{eff}$) of the symmetric mode, the green solid line is for Re($n_{eff}$) of the anti-symmetric mode, the dashed red line is the figure of merit of the symmetric mode, and the light blue dashed line is the figure of merit of the anti-symmetric mode.

Due to the extremely large wavenumber supported by this configuration, additional modes with high order oscillations are found in our calculation. If $d$ is not too small, e.g. 20 nm, only two modes are supported. However, if the two ribbons get closer to each other, wavenumber $k$ will grow significantly. Then more oscillations along the lateral direction will occur, resulting in more waveguide modes. For example, when $d$ is 2 nm, one additional mode appears (Fig. 9). This characteristic is potentially useful for the design of e.g. ultra-compact MMI (multimode interferometer) devices.

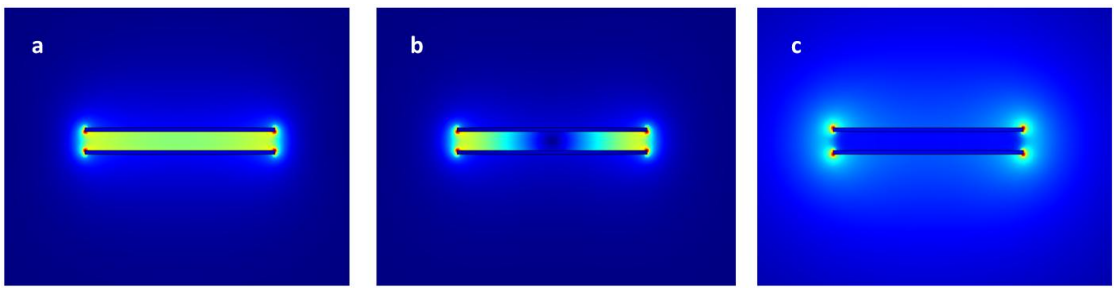

**Figure 9 | The modes when $d$ is small enough (2 nm).** (**a**) and (**c**) are for the mode (which originally is the symmetric mode) that is formed through the coupling/hybridization of the two ribbons. (**b**) is a high order mode in the x direction.

Since a single freestanding graphene ribbon can support a plasmonic waveguide mode with an extremely large wave number, this configuration has high potential for design of some ultra-compact optical devices. Here as an example we give a nano-ring cavity based on a single freestanding graphene ribbon. In our design, the width of the graphene ribbon is chosen to be 20 nm. It is well known that a round-trip phase accumulation of integral multiple of $2\pi$ can result in a resonant mode. With an inner radius $r = 38$ nm, we see a resonant peak at 30 THz (corresponding to a vacuum wavelength of about 10 μm) in our numerical simulation with CST software. The mode profile shows that it is a 4th-order ring cavity mode. The *Q* value ($Q = f /\Delta f$, *f* is the frequency, $\Delta f$ is the linewidth) is 42.3. Considering the ring size is on the order of nanometer, the cavity size (compared to the vacuum wavelength) is surprisingly small.

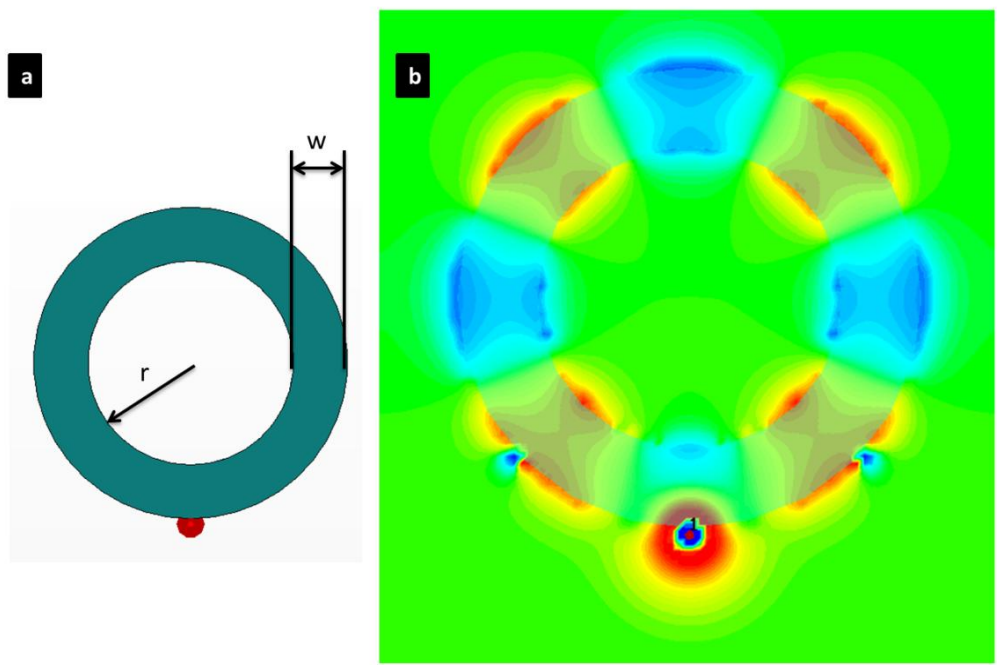

**Figure 10 | The structure and electric field distribution of the ring cavity.** (**a**) The proposed ring cavity with graphene ribbon width *w* = 20 nm and inner radius *r* = 38 nm. (**b**) The electric field distribution of the cavity mode at f = 30 THz, which is excited by a discrete port in CST.

In summary, the unique properties of plasmonic waveguides based on graphene nano-ribbons have been investigated. The guided modes are tightly confined in both the lateral direction and the propagation direction. A single mode operation region has been identified if the ribbon width is small enough. Due to the tight confinement, a small mode area and high effective refractive index can be achieved. The low-loss waveguide structure with an embedded low index silica layer between the graphene layer and the silicon substrate has been proposed to reduce the propagation loss and increase the FOM of the plasmonic waveguide. The coupled configurations with two identical graphene ribbons also exhibit interesting properties. In particular, the side-side-coupling can further reduce the waveguide mode area, while the top-bottom-coupling can result in much larger effective indices than an isolated graphene ribbon. A nano-ring cavity of extremely small size based on a graphene ribbon waveguide has been shown to support a cavity resonance at far infrared range.